# Customizing ChatGPT for Second Language Speaking Practice: Genuine Support or Just a Marketing Gimmick?

Fanfei Meng, University of Pennsylvania, fanfei@upenn.edu

**Abstract:** ChatGPT, with its customization features and Voice Mode, has the potential for more engaging and personalized ESL (English as a second language) education. This study examines the efficacy of customized ChatGPT conversational features in facilitating ESL speaking practice, comparing the performance of four versions of ChatGPT Voice Mode: uncustomized *Standard* mode, uncustomized *Advanced* mode, customized *Standard* mode, and customized *Advanced* mode. Customization was guided by prompt engineering principles and grounded in relevant theories, including Motivation Theory, Culturally Responsive Teaching (CRT), Communicative Language Teaching (CLT), and the Affective Filter Hypothesis. Content analysis found that customized versions generally provided more balanced feedback and emotional support, contributing to a positive and motivating learning environment. However, cultural responsiveness did not show significant improvement despite targeted customization efforts. These initial findings suggest that customization could enhance ChatGPT's capacity as a more effective language tutor, with the standard model already capable of meeting the learning needs. The study underscores the importance of prompt engineering and AI literacy in maximizing AI's potential in language learning.

## Introduction

The rapid development of generative AI tools, such as ChatGPT, has unlocked significant potential in education, especially in creating virtual teaching assistants that offer personalized learning experiences (Garrido-Merchán et al., 2024). A systematic review of Artificial Intelligence in Education (AIE) highlights language learning as the most frequently researched subject area (Crompton & Burke, 2023), driven by AI tool's ability to provide personalized learning experiences, immediate feedback, and a low-pressure environment conducive to language acquisition (Hegazy, 2024). Most studies have focused on the impact of general AI on reading, writing, and vocabulary acquisition (Liang et al., 2021), demonstrating benefits such as improved reading comprehension (Jiang, 2022; Sudin, 2024), facilitated vocabulary learning with reduced anxiety (Hsu, Chang, & Jen, 2024; Liang & Zhang, 2024), and enhanced academic writing and learner motivation (Song & Song, 2024; Marzuki et al., 2023). However, comparatively less attention has been given to leveraging AI for developing speaking skills, highlighting the need to explore the potential of large language models (LLMs) in supporting speaking practices in English as a Second Language (ESL) education.

In response to the growing demand for more personalized user experiences, OpenAI introduced the *Custom Instructions* feature for ChatGPT in July 2023, allowing users to specify preferences and requirements to better tailor AI's responses to individual needs. By July 2024, OpenAI further expanded ChatGPT's capabilities with *Advanced* Voice Mode, which allows real-time conversational interaction via voice input and output. This mode leverages GPT-4o's multimodal capabilities to facilitate natural, emotionally responsive spoken conversations, surpassing the earlier *Standard* Voice Mode that required separate transcription processes before generating responses (OpenAI, n.d.). These features promise more engaging and personalized language learning experiences.

Despite their potential, studies on the educational application of customized AI remain limited. A preliminary study by Garrido-Merchán et al. (2024) compared the performances of a customized GPT model for business statistics learning with GPT-4 Turbo, revealing no significant differences in outputs. Yet, research into the application of customization specifically for language education, particularly for ESL speaking practice, remains underexplored.

Recognizing the importance of *prompt engineering* as a vital technique for optimizing LLM performance in educational contexts (Ye et al., 2023; Liu et al., 2021), Hegazy (2024) further emphasizes the importance of integrating language learning theories into prompt design to ensure both linguistic accuracy and pedagogical value in the task of generating teaching reading materials. However, there has been little effort to systematically combine language learning theories, learning science principles, and prompting strategies to customize AI for language learning remains largely unexplored.

To address these gaps, this study investigates the potential of customized GPT voice modes in facilitating ESL speaking practice. The customization process is grounded in four foundational theories of learning science and ESL education. These include Motivation Theory, positing that external incentives trigger internal motives, driving actions and shaping future motivation through their outcomes (Urhahne & Wijnia, 2023), supporting prompt design that sustains learner engagement; Culturally Responsive Teaching (CRT), which incorporates students' cultural backgrounds into education to enhance their academic and personal growth (Gay, 2013), informing culturally sensitive AI customization; Communicative Language Teaching (CLT), which guides the customization of LLMs to promote authentic communication and meaningful language use (Qasserras, 2023); and the Affective Filter Hypothesis, which helps in lowering learner anxiety and fostering a supportive learning environment conducive to language acquisition (Du, 2009). By integrating these theories with prompting strategies, this study aims to bridge the gaps in existing research on customizing AI for ESL education, with a specific focus on speaking practice.

Building on this theoretical foundation, the study conducts a comparative analysis of four distinct configurations of GPT voice modes to evaluate their effectiveness in facilitating ESL speaking practice. Configurations include (1) *uncustomized Standard* Voice Mode, (2) *customized Standard* Voice Mode, (3) *uncustomized Advanced* Voice Mode, and (4) *customized Advanced* Voice Mode. Each mode is tested as an ESL speaking practice partner to assess differences in performance between *customized* and *uncustomized* versions, as well as between *Standard* and *Advanced* Voice Modes. This structured comparison examines how both customization and capability levels (standard vs. advanced) influence the AI's performance. The objective of this pilot study is to offer initial insights into the potential value of ChatGPT's customized voice functions for enhancing ESL learning experiences, contributing to the early understanding of the role of customized AI in education contexts.

### Research questions

- What is the impact of prompts informed by educational theories on the performance of GPT voice modes as ESL speaking practice partners?
- To what extent does customization improve GPT's effectiveness in facilitating ESL speaking practice compared to non-customized versions?
- How do the capability levels of AI models (standard vs. advanced voice modes) influence their effectiveness as ESL-speaking tutors?

## Methods

This preliminary study aims to evaluate the potential of customizing GPT in enhancing their role as supportive tutors for second language (L2) speaking practice. The researcher of this study is a master's student in the Learning Sciences and Technologies program at a university in the northeastern United States. Four versions of conversation models within ChatGPT were comparatively analyzed to explore performance differences: *uncustomized Standard* Voice Mode, *uncustomized Advanced* Voice Mode, *customized Standard* Voice Mode, and *customized Advanced* Voice Mode.

### Prompt design for customization

The customization of the models adhered to the *Template-Based Prompting* strategy (Kung et al., 2022; Liu et al., 2021; Wang et al., 2022), using prompt configuration to define AI behavior through detailed textual inputs without modifying its underlying architecture or parameters (Figure 1). This approach to prompt configuration ensures accessibility for students and teachers in educational settings with limited technical resources. The prompt provided to each customized model was identical, adhering to the same structure and incorporating three main components: defining the AI's role as a language tutor, specifying the desired tone and response style, and structuring the interaction procedure based on the IELTS speaking exam format. This standardization ensured consistency across all customized models and facilitated a fair comparison of their responses.

Additionally, key elements from Motivation Theory, Culturally Responsive Teaching (CRT), Communicative Language Teaching (CLT), and the Affective Filter Hypothesis were integrated into the prompt, ensuring the AI's responses were effective in supporting language learning (Hegazy, 2024). Motivation Theory informed the use of encouragement and praise language. The Affective Filter Hypothesis guided the AI to adopt a supportive, anxiety-reducing tone. CRT was reflected by relating the conversation to the learner's background as revealed in their responses, and CLT shaped the use of open-ended, conversational prompts to simulate authentic interaction.

To ensure uniformity, the two uncustomized models are also asked to follow the IELTS conversation procedure during interactions. Identical short-answer inputs were provided to all models to facilitate fair comparison. Similar inputs were provided to all models during the initial and final interaction segments, and for the IELTS long-turn segment (Part 2), the same responses were used to ensure consistency in input across all models (Figure 2).

**Figure 1**
*Prompt Configuration*

**Figure 2**
*Transcripts of the Interactions with the Four Modes*

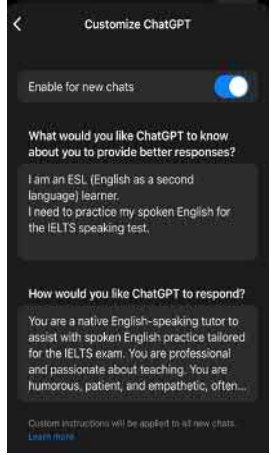
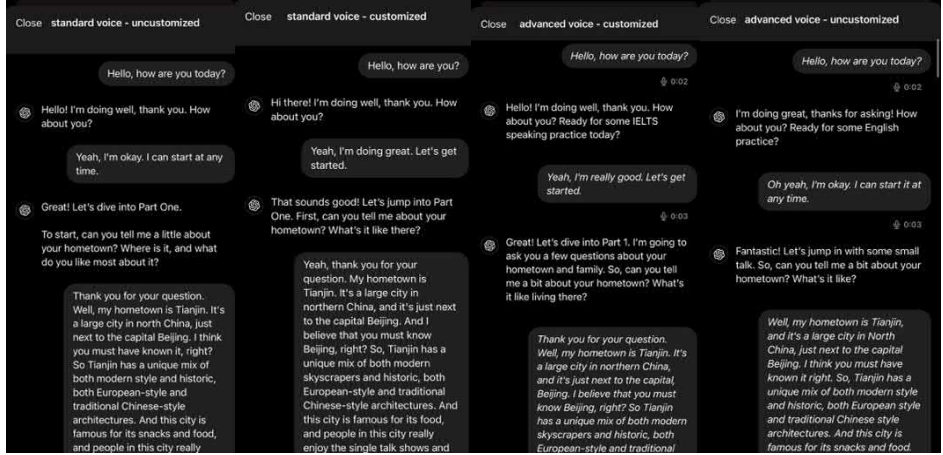

## Data analysis

The responses collected from the four models were analyzed using content analysis (Hsieh & Shannon, 2005). The analysis began with an inductive open coding phase to identify emergent characteristics directly from the data, allowing for an unbiased exploration of patterns and themes. Following this, the initial codes were deductively aligned with the four theories used in the customization prompt design. This step refined the codes, either categorizing them into established codes or generating new ones that better represented the data. Each sentence was treated as the unit of analysis for detailed coding. A subset (20%) was double coded by the researcher, yielding substantial agreement ($\kappa=0.78$). Code frequencies were recorded to highlight the most prevalent response characteristics across the models.

## Results

Fourteen codes were identified and organized into five main categories: content response, cultural responsiveness, emotional support, corrective feedback, and social interaction. The proportion of each category to the overall response was calculated based on code frequency (Figure 3). The results demonstrate notable differences between the customized and uncustomized modes, with customized models generally providing a more balanced and comprehensive set of responses. The uncustomized modes, on the other hand, leaned more heavily towards information-focused responses with less engagement in other supportive aspects.

**Figure 3**
*Proportional Distribution of Response Categories Across the Four Modes*

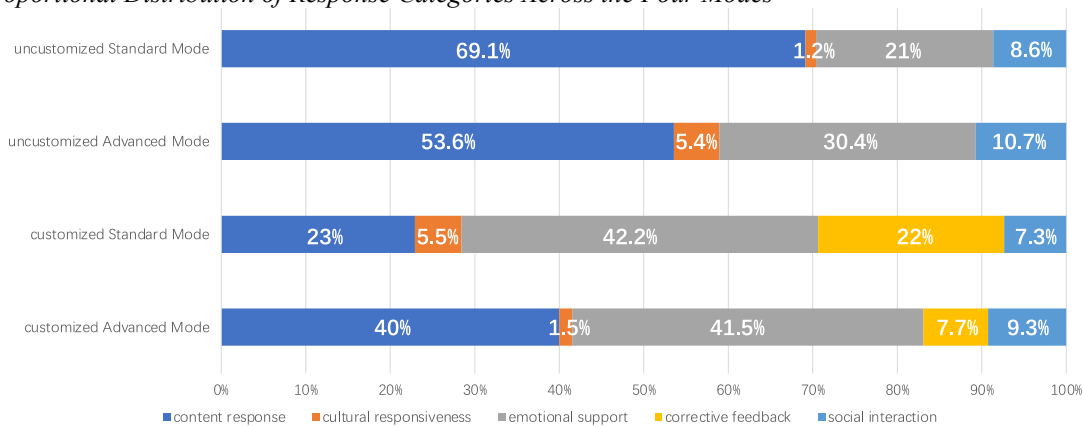

## Content response

The content response category includes reflective summarization, actionable guidance, and inference, focusing on providing relevant information based on user input. This was the largest segment across all four models, showing all modes excel in delivering substantial information, such as summarizing learner's ideas or inferring intentions. The *uncustomized Standard* Mode had the highest proportion of content responses at 69.1%, which decreased significantly to 40% in *Advanced* Mode and 23% in *Standard* Mode after customization. This shift indicates that customization enables GPT to balance informational responses with other supportive features, promoting a broader range of outputs essential for effective language practice.

### Emotional support

Emotional support, including content appreciation, anxiety reduction, positive reinforcement, and engagement, was more prevalent in the *customized* modes, suggesting the successful integration of the Motivation Theory and the Affective Filter Hypothesis into customized prompt. The *customized Standard* Mode had the highest proportion at 42.2%, followed closely by the *customized Advanced* Mode at 41.5%. This indicates that these customized models effectively provided more emotionally supportive responses than the uncustomized ones, highlighting the potential of customization to create a more engaging and motivating experience.

### Corrective feedback

Corrective feedback is essential in language learning as it helps learners consciously identify and correct errors, reinforcing accurate language use and promoting long-term language development (Ellis, 2021). Feedback and error correction related to issues in fluency, grammar, vocabulary, and pronunciation were present in the customized modes' output, showing significant differences compared to the uncustomized models. Notably, the *customized Standard* Mode stood out with feedback making up 22% of its total responses, while the customized *Advanced* Mode had a feedback proportion of 7.7%. In contrast, GPT models without customization lack the agency to provide error correction to learners in facilitating language learning.

### Social interaction and cultural responsiveness

Social interaction refers to conversational exchanges that foster interpersonal connections. These types of interactions can help develop a learner's communicative competence to use appropriate language in social contexts beyond just linguistic skills. The data results for social interaction show no significant differences among the four models, suggesting that all models inherently possess characteristics that support social interaction, but customization does not influence their performance in this category.

A similar trend is observed in the cultural responsiveness category, which involves cultural awareness and adaptation in response generation. Despite the integration of CRT principles in the customized prompts, their impact on the models' responses was limited. Notably, the customized Advanced Mode showed the second-lowest CRT rating, highlighting a potential limitation in customization and suggesting a need for further study. Future improvements may require integrating externally sourced, culturally relevant prompts to enhance responsiveness.

## Discussion and implication

The results of this study provide initial evidence of the educational value of customizing GPT models for ESL speaking practice. However, as a single-case pilot, broader data collection is needed to assess the consistency and generalizability of these findings. In this study, customized models showed notable advantages over uncustomized ones, especially in delivering feedback and emotional support, thereby fostering a more positive learning environment. These findings suggest that theory-informed customization can enhance AI's educational impact by better addressing learner needs. Notably, efforts to embed cultural responsiveness through customization yielded limited results, indicating the need for more effective prompting strategies. This highlights a key concern for educators—the potential limitations of AI in adapting to learners' cultural contexts.

Regarding the third research question, results showed that the advanced model did not significantly outperform the standard one in core areas like feedback and emotional support. This suggests that the standard model may already meet essential needs, making it a more cost-effective option for educational settings. Still, the advanced version's more human-like tone and faster response times may be valuable areas for future exploration.

This study also reinforces the importance of prompt engineering in shaping AI output qualities. While AI shows great promise for personalized learning, its effectiveness is heavily dependent on how it is prompted. As AI tools become more accessible, improving AI literacy among educators and learners is crucial to fully realize their educational potential. Additionally, as AI grows more emotionally expressive, future research should consider the ethical implications of simulating interpersonal relationships in language learning.